# Electron-hole pair condensation in Graphene/MoS$_2$ heterointerface

Min-Kyu Joo[1,2†,*], Youngjo Jin[1,2†], Byoung Hee Moon[1], Hyun Kim[1,2], Sanghyub Lee[1,2] & Young Hee Lee[1,2,3*]

[1]*Center for Integrated Nanostructure Physics (CINAP), Institute for Basic Science (IBS), Suwon 16419, Republic of Korea*

[2]*Department of Energy Science, Sungkyunkwan University (SKKU), Suwon 16419, Republic of Korea*

[3]*Department of Physics, Sungkyunkwan University (SKKU), Suwon 16419, Republic of Korea*

[*]Correspondence to: mkjoo@skku.edu (M.-K. Joo) and leeyoung@skku.edu (Y.H. Lee)

[†]These authors contributed equally to this work.

Excitons are electron-hole (*e-h*) pair quasiparticles, which may form a Bose-Einstein condensate (BEC) and collapse into the phase coherent state at low temperature[1-3]. However, because of ephemeral strength of pairing, a clear evidence for BEC in electron-hole system has not yet been observed. Here, we report electron-hole pair condensation in graphene (Gr)/MoS$_2$ heterointerface at 10K without magnetic field. As a direct indication of *e-h* pair condensation[3-6], we demonstrate a vanished Hall drag voltage and the resultant divergence of drag resistance. While strong excitons are formed at Gr/MoS$_2$ heterointerface without insulating layer, carrier recombination *via* interlayer tunneling of carriers is suppressed by the vertical *p*-Gr/*n*-MoS$_2$ junction barrier, consequently yielding high BEC temperature of 10K, ~1000 times higher than that of two-dimensional electron gas in III-V quantum wells[4,7-8]. The observed excitonic transport is mainly governed by the interfacial properties of the Gr/MoS$_2$ heterostructure, rather than the intrinsic properties of each layer. Our approach with available large-area monolayer graphene and MoS$_2$ provides a high feasibility for quantum dissipationless electronics towards integration.

Dissipationless transport in solids as exemplified superfluidity or superconductivity is accomplished through the condensation of bosonic states, called a Bose-Einstein condensate (BEC)[2,9]. Although a theory predicts that excitonic bound state of electron-hole (*e-h*) pairs can be observed near room temperature in two-dimensional layered materials due to strong exciton binding energy[10-11], realizing *e-h* pair condensation has been challenging even at low cryogenic temperature because excitonic bosons easily recombine or deviate from coherent superfluidic state[2,3,9].

A simple way to suppress exciton recombination and provide a sufficiently short distance for strong Coulomb interaction is to squeeze a thin insulating layer between two parallel conducting layers, so called 'Coulomb drag configuration'[2-3,12]. In conventional methods, an excitonic BEC has been implemented in two-dimensional electron gases (2DEG) of GaAs double quantum wells[4,7,8]. At the fractional quantum Hall state, the double electron system becomes equivalent to electron-hole system by making a particle-hole transformation in one of the two quantum wells[2]. The degree of interaction is estimated by the parameter *d/l*, where *d* is separation distance between active and passive layers, and $l$ $(=(\hbar/eB)^{0.5})$ is magnetic length (here, $\hbar$ is Planck constant divided by $2\pi$, *e* is electron charge, and *B* is magnetic field)[13-14]. The signature for the formation of excitonic BEC in a strong coupling regime $d/l \leq 2$ is reentrant quantized Hall state in the normal configuration[4,8,14-15] or vanished Hall voltage in the counterflow configuration at the total Landau level ($v_T$) equal to one[2,4,8,15]. Owing to the advanced technology, the structure of electron-hole double quantum wells could be constructed but the distinct evidence for BEC in the absence of magnetic fields of the vanished Hall voltage and resultant divergence of drag resistance has never been provided yet due to the rather weak Coulomb interaction between electrons and holes *via* thick insulating layer (few tens of nanometer), although this barrier is essential in sufficiently suppressing recombination through interlayer tunneling and ruling out disorder effects in the system[3,6,16].

Recently, similar experiments with graphene (Gr), where mono- and bi-layer Gr double-gate structures, separated by a few layered hexagonal boron nitride (*h*-BN), allow for strong coupling regime and demonstrate remarkable progresses in Coulomb drag as well as excitonic superfluidity experiments[3,14-15,17-18]. These atomically thin layered materials leave for active channel (current-driving) and passive channel (drag-voltage-measuring) layers to be quantized and simultaneously to isolate both layers. Moreover, unlike GaAs double quantum wells, the beneficial features of *h*-BN with a few nanometer thickness and low electrical permittivity ($\varepsilon \sim$ 4.0) provide a strongly coupled regime[15]. Besides, the wide range of carrier-type and -density tunability in graphene helps to reveal the new Coulomb drag mechanisms, such as anomalously strong drag[4], negative drag[18] at double charge neutrality points of graphene, and excitonic superfluidity[14-15]. However, any traces of the formation of exciton condensation in electron-hole system without magnetic fields are yet to be reported.

In this work, we demonstrate two exclusive features of BEC; i) a vanished Hall drag voltage and ii) the resultant divergence of drag resistance by constructing simple monolayer Gr/monolayer $MoS_2$ heterointerface (see methods and supplementary note 1). Each material is prepared by chemical vapor deposition (CVD). Since both layers are in contact *via* van der Waals gap without insulating layer, an extremely strong coupling limit ($d/l \approx 0.045$ where $d = 3.1$ Å when $B = 14$ T) is achieved. Yet, the vertical *p*-Gr/*n*-$MoS_2$ junction barrier exists and consequently suppresses carrier recombination *via* tunneling. This enhances *e-h* indirect exciton stability and increases the critical temperature for *e-h* exciton condensation.

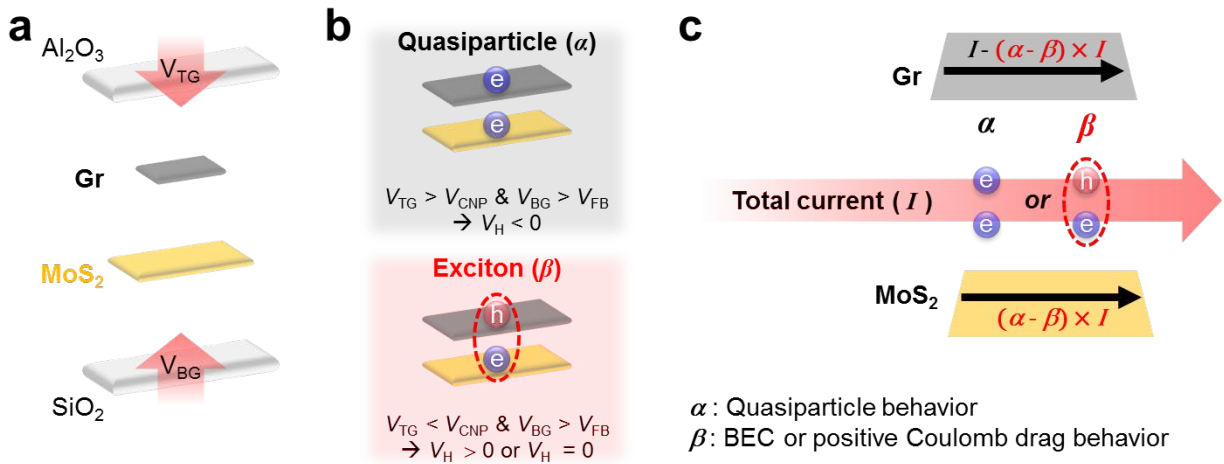

**Figure 1 | Exciton pairs in Gr/MoS$_2$ heterostructures. a,** Schematic for Gr/MoS$_2$ heterostructure with dual-gate symmetry. Gr is isolated deliberately from source and drain electrodes. **b,** Carrier transport mechanism under different top- and bottom-gate bias conditions. Exciton pairs *via* the Coulomb interaction are formed when; Gr and MoS$_2$, are, *p*-type and *n*-type, respectively (bottom panel); while quasiparticle transport, governed by Boltzmann transport, is

expected when both Gr and MoS$_2$ are *n*-type. **c,** Total current contribution of the two factors, $\alpha$ and $\beta$, representing the portion of quasiparticle behavior and exciton condensation (or Coulomb drag), respectively.

Figure 1a illustrates double-gated Gr/MoS$_2$ heterostructures to investigate two representative conduction regimes; *i*) *p*-type Gr on *n*-type MoS$_2$ and *ii*) *n*-type Gr on *n*-type MoS$_2$. Provided that the back-gate bias ($V_{BG}$) exceeds a flat-band voltage ($V_{FB}$) of bottom MoS$_2$, guaranteeing *n*-type MoS$_2$, *e-h* indirect exciton pairs will be created when the top-gate bias ($V_{TG}$) is lower than the charge neutrality point ($V_{CNP}$) of Gr. Exciton recombination *via* undesired tunneling is largely suppressed by the *p-n* junction barrier at the heterointerface on the basis of energy band diagram (supplementary note 2 and 3). This gives rise to a counterflow current ($I_{CF}$) in MoS$_2$ layer in Gr/MoS$_2$ heterostructures because the channel resistance of MoS$_2$ is relatively larger than that of Gr in regime *i*. Conversely, when $V_{TG} > V_{CNP}$ where the graphene is *n*-type, the quasiparticle behavior is expected to be governed by Boltzmann transport.

To differentiate this complex transport phenomenon, $\alpha$ and $\beta$ factors are introduced here (Figure 1b). $\alpha$ indicates the quasiparticle behavior and $\beta$ stands for *e-h* pair condensation or Coulomb drag effects with a total current *I*. For example, when excitonic bosons and/or Coulomb drag effects are dominant, $\beta$ is greater than $\alpha$, and thus the Hall voltage ($V_H$) should be vanished (*e-h* pair condensation) or larger than zero (indicating *p*-type transport due to strong Coulomb drag effect). In addition to $V_H$, strong *e-h* Coulomb interaction also leads to a positive longitudinal potential drop ($V_{XX}$, obtained from 4-probe configuration) in Gr, but negative $V_{XX}$ in MoS$_2$ because of $I_{CF}$. Meanwhile, when *e-h* Coulomb drag effects are negligible, $\alpha$ should be larger than $\beta$, resulting in $V_H$ lower than zero (indicating *n*-type transport) and positive $V_{XX}$ in both Gr and MoS$_2$ (Figure 1c).

We first examine the existence of $I_{CF}$ in Gr/MoS$_2$ heterostructures by constructing $V_{XX}$ probes separately in Gr ($V_{XX\_Gr}$) and MoS$_2$ ($V_{XX\_MoS_2}$) (Figure 2a). The room-temperature transfer curves, for Gr/MoS$_2$ heterostructure, isolated Gr, and isolated MoS$_2$ on the same wafer, are plotted together in Figure 2b. The on-current of Gr/MoS$_2$ heterostructure is roughly increased by one order of magnitude compared to that of isolated MoS$_2$, while maintaining similar or even larger on/off current ratio. The flat-band voltage of Gr/MoS$_2$ device is downshifted compared to that of MoS$_2$, due to the passivated graphene over layer. The maximum four-probe mobility ($\mu_{FE\_4p}$) in Gr/MoS$_2$ heterostructure reaches 12,000 cm$^2$V$^{-1}$s$^{-1}$ (Figure 2c), which is doubled from that of pristine Gr, implying the importance of heterostructure.

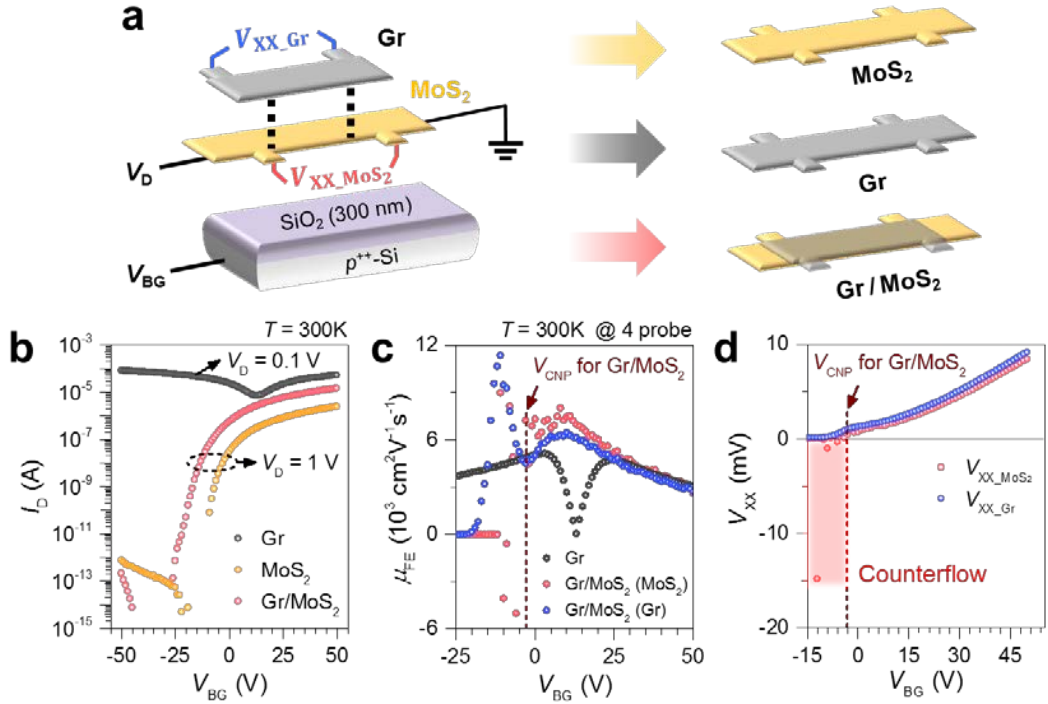

**Figure 2 | Counterflow current ($I_{CF}$) in Gr/MoS$_2$ heterostructures. a,** Device schematic for the $I_{CF}$ measurements. Three devices (MoS$_2$, graphene, Gr/MoS$_2$) were fabricated on the same wafer. To identify $I_{CF}$, the longitudinal potential drops in Gr ($V_{XX\_Gr}$) and MoS$_2$ ($V_{XX\_MoS_2}$) were separately measured, while applying the drain current through the MoS$_2$ contact. **b,** Room-temperature transfer curves ($I_D$−$V_{BG}$) of Gr/MoS$_2$ heterostructure (red), MoS$_2$ (orange), and graphene (black). **c,** Four-probe field-effect mobility ($\mu_{FE\_4p}$) measured at the bottom-MoS$_2$ (red) and top-Gr (blue). Gr (black) was separately measured. **d,** $V_{XX}$ of the Gr/MoS$_2$ heterostructure device was measured at the bottom-MoS$_2$ (red) and top-graphene (blue). Negative $V_{XX\_MoS_2}$ indicates the existence of $I_{CF}$ in the MoS$_2$ layer.

In general, $\mu_{FE\_4p}$ measured from Gr and MoS$_2$ should be similar, if the Gr/MoS$_2$ heterostructure is considered as a one-body system. In our Gr/MoS$_2$ heterostructure, the charge transfer occurs from MoS$_2$ to Gr, and thus $V_{CNP}$ of graphene in Gr/MoS$_2$ heterostructure is downshifted to ~ 0 V, compared to $V_{CNP}$ of isolated Gr (~18 V). Consequently, the vertical *p-n* junction barrier is established; when graphene is *p*-type ($V_{BG} \leq V_{CNP}$) while MoS$_2$ is still *n*-type for Gr/MoS$_2$ heterostructure. The formation of *p-n* junction barrier suppresses the tunneling current, preventing exciton recombination. In this sense, our Gr/MoS$_2$ heterostructure is no longer considered as a one-body system, opening the possibility of observing $I_{CF}$. Such $I_{CF}$ is monitored in negative $V_{XX\_MoS_2}$ and positive $V_{XX\_Gr}$ at $V_{BG} \leq V_{CNP}$ (Figure 2d). Although the range of $I_{CF}$ is narrow and fluctuates due to thermal effects at room temperature, $I_{CF}$ allows the Gr/MoS$_2$ heterostructure to mimic Coulomb drag configuration.

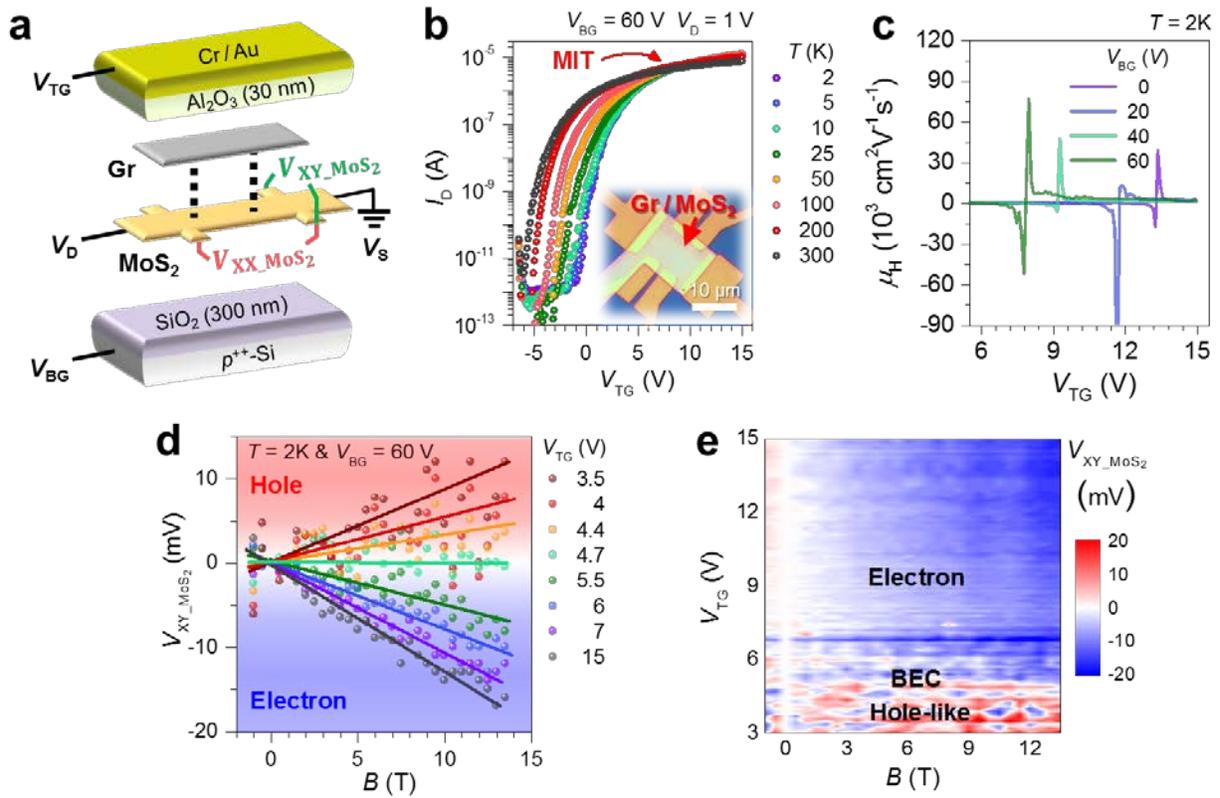

**Figure 3 | Vanished Hall drag voltage in Gr/MoS$_2$ heterostructures. a,** Double-gated Gr/MoS$_2$ heterostructures with a 6-probe Hall-bar structure. The longitudinal ($V_{XX\_MoS_2}$) and transversal ($V_{XY\_MoS_2}$) potential drops were measured in the MoS$_2$ layer. **b,** *T*-dependent transfer curves ($I_D-V_{TG}$) of Gr/MoS$_2$ heterostructure at $V_{BG} = 60$ V and $V_D = 1$ V. (Inset: Optical image of the device) **c,** $V_{TG}$-dependent $\mu_H$ as a function of $V_{BG}$ at $T = 2$K. **d,** *B*-dependent $V_{XY\_MoS_2}$, for different $V_{TG}$ at $T = 2$K for $V_{BG} = 60$ V. **e,** Corresponding 2D contour plot for $V_{XY\_MoS_2}$. Sign reversal of $V_{XY\_MoS_2}$ is observed at $V_{TG} \sim 4.7$ V.

To control a wide range of carrier-type and -density in each layer, double-gated Gr/MoS$_2$ heterostructures were fabricated by a standard wet transfer method. The device schematic (Figure 3a) and optical image (inset of Figure 3b) are shown, respectively. The top Gr layer is not physically contacted with source and drain electrodes. When the total drain current ($I_D$) flows along the heterostructure, the potential drops along the longitudinal ($V_{XX\_MoS_2}$) and transverse ($V_{XY\_MoS_2}$) directions, are separately measured in MoS$_2$ layer. Figure 3b presents the temperature ($T$)-dependent transfer curves of Gr/MoS$_2$ heterostructure as a function of $V_{TG}$ at a fixed drain voltage $V_D = 1$ V and $V_{BG} = 60$ V for *n*-type conduction in MoS$_2$. A clear metal-insulator transition (MIT) is observed at $V_{TG} \sim 7.5$ V, implying a clean interfacial quality, because the MIT is largely disturbed by oxide traps, defects, and disorders at various interfaces of Gr/MoS$_2$[19]. The obtained maximum Hall mobility ($\mu_H = (L/W) V_{XY\_MoS_2} V_{XX\_MoS_2}^{-1} B^{-1}$, where $L/W$ are channel aspect ratio for 4-probe configuration) is 76,600 cm$^2$V$^{-1}$s$^{-1}$ when $T = 2$K and $V_{BG} = 60$ V (Figure 3c), outperforming carrier mobility of individual Gr and MoS$_2$ layer.

As a prominent signature of *e-h* pair condensation, we present a vanished $V_{XY\_MoS_2}$, relative to the current flow, as a function of $B$[5,6,20]. In general, the Lorentz force pushes carriers vertically inside MoS$_2$ layer, producing a finite Hall voltage difference. According to a theory[2-3,6], however, excitonic bosons follow dissipationless transport in the presence of $I_{CF}$, resulting in zero values of $V_{XY\_MoS_2}$ and $V_{XX\_MoS_2}$ in Gr/MoS$_2$ when the Lorentz force and *e-h* binding energy are well balanced. Figure 3d displays $B$-dependent $V_{XY\_MoS_2}$ at various $V_{TG}$ with $T = 2$K. The $V_{XY\_MoS_2}$ slope varies from positive (hole-like) to negative (electron) as $V_{TG}$ increases. The crossover ($V_{XY\_MoS_2} \sim 0$) occurs at $V_{TG} \sim 4.7$ V independent of $B$, indicating the existence of *e-h* pair condensation. Figure 3e shows the corresponding 2D contour plot of $V_{XY\_MoS_2}$. The alternating crossover (indicated by white) is attributed to the formation of Landau levels. When $V_{TG}$ is less than the *e-h* pair condensation ($V_{TG} \sim 4.7$ V), electrons in MoS$_2$ are suppressed, while holes in Gr are enriched, resulting in an imbalance between the electron and hole populations. Therefore, the total current from MoS$_2$ contact is preferred to mostly flow along Gr, generating $I_{CF}$ along MoS$_2$, (holes in graphene drag the electrons in MoS$_2$, exhibiting hole-like behavior), leading to positive $V_{XY\_MoS_2}$, mainly *via* Coulomb drag force. Meanwhile, the electron carrier densities populate both layers at $V_{TG} > 4.7$ V (electrons in graphene drag electrons in MoS$_2$), leading to a negative $V_{XY\_MoS_2}$ (see supplementary note 3).

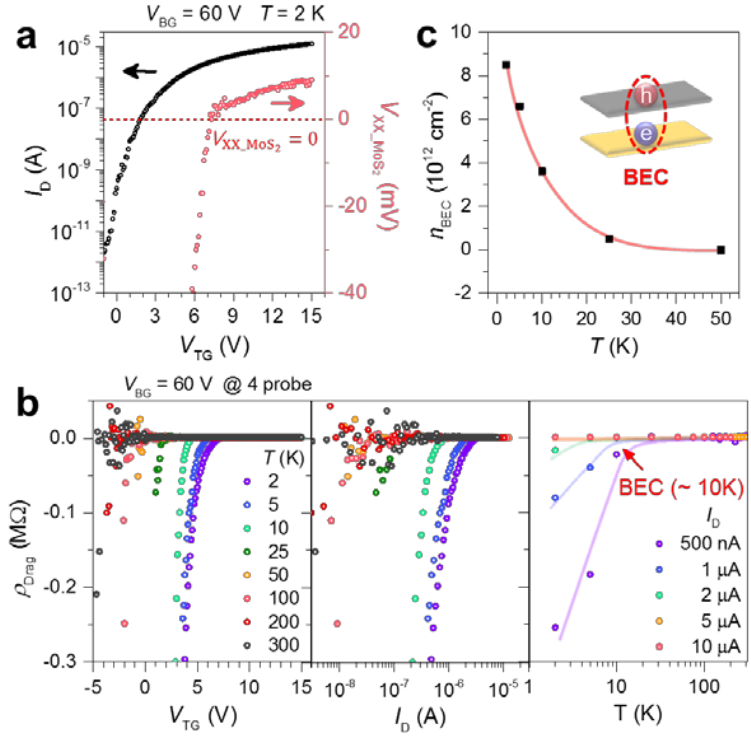

**Figure 4 | Exciton condensation in Gr/MoS$_2$ heterostructures. a,** $V_{TG}$-dependent $V_{XX\_MoS_2}$ (red) and $I_D$ (black) at $T = 2K$. The polarity reversal of $V_{XX\_MoS_2}$ is occurred at $V_{TG} = 7.5$ V. **b,** The following curves are shown; $V_{TG}$-dependent $\rho_{Drag}$ (=$V_{XX\_MoS_2}/I_D$) (left panel); $I_D$-dependent $\rho_{Drag}$ (middle panel) at various $T$ for $V_{BG} = 60$ V; and, $T$-dependent $\rho_{Drag}$ for selected $I_D = 500$ nA, 1 μA, 2 μA, 5 μA, 10 μA (right panel). **c,** $T$-dependent $n_{BEC}$ (carrier density for BEC).

Figure 4a displays the $V_{TG}$-dependent $V_{XX\_MoS_2}$ at 2K. Interestingly, $V_{XX\_MoS_2}$ changes the polarity at $V_{TG}$ (or $V_{CNP}$) ~ 7.5 V, which is seemingly similar to the MIT point. In other words, the metallic phase appears at $V_{TG} > 7.5$ V, where the Gr/MoS$_2$ heterostructure can be considered as a one-body system, which is also seen in Figure 2c ($V_{BG} - V_{CNP} > 0$ V). Meanwhile, a wide range of negative $V_{XX\_MoS_2}$ is clearly observed with $I_{CF}$, when $V_{TG} < 7.5$ V. $V_{XX\_MoS_2}$ is zero at $V_{TG}$ ~ 7.5 V, which can be interpreted as the balanced state between positive Coulomb drag including BEC (negative sign) and quasiparticle contribution (positive sign). The $V_{TG}$ values for zero $V_{XY\_MoS_2}$ and $V_{XX\_MoS_2}$ differ slightly. This can be understood by considering the remaining free carriers, which do not form excitons in the heterointerface channel. Besides, the perfect $I_{CF}$ circumstances, wherein the same amplitude of current in both layers but opposite direction, cannot be achieved in our current system, which therefore leads to different $V_{TG}$ positions, for zero $V_{XY\_MoS_2}$ and $V_{XX\_MoS_2}$.

To further differentiate *e-h* pair condensation from positive Coulomb drag, the *T*-dependent Coulomb drag resistance ($\rho_{Drag} = V_{XX\_MoS_2}/I_D$) is plotted as a function of $V_{TG}$ and $I_D$ in Figure 4b. For a positive Coulomb drag, where $\rho_{Drag}$ is dominated by the momentum and energy transfer between two layers, $\rho_{Drag}$ is proportional to $T^2$ at $T \leq T_F$ ($T_F$ is the Fermi-temperature) and vanishes at $T = 0$[12,18,21-22]. In our case, however, the divergence of $\rho_{Drag}$ ($= I_{CF}R_{MoS_2}/I_D$) becomes more distinct when *T* decreases for $V_{TG} \leq V_{CNP}$. Owing to the discrepancy between $I_{CF}$ and $I_D$, percolative *e-h* condensation puddles with the resistivity of MoS$_2$ passive layer induce a finite resistance of $R_{MoS_2}$. Meanwhile, $I_{CF}$ increases with decreasing *T* due to enhanced interlayer Coulomb interaction, giving rise to the divergence of $\rho_{Drag}$. In addition, the divergence of $\rho_{Drag}$ is drastically suppressed at $V_{TG} \geq V_{CNP}$, which is attributed to the enhanced exciton recombination *via* interlayer tunneling through the lowered *p-n* junction barrier height. This is consistent with theoretical prediction for exciton condensation rather than Coulomb drag[6,20].

To demonstrate *T*-dependent $\rho_{Drag}$, we selected several $I_D$ from the middle panel of Figure 4b, and plotted them in the right panel. When $I_D$ is less than 2 μA, $\rho_{Drag}$ diverges nearby $T \sim 10$K and approaches the resistance of bottom MoS$_2$ layer, which is the theoretically predicted signature for exciton condensation[2-3,6]. In general, exciton condensation occurs at low carrier density and low temperature. However, in our case, BEC occurs even at higher carrier density when $T < 10$ K (see Figure 4c and supplementary note 4). This provides room to obtain an even higher condensation temperature *via* the modulation of carrier density. Consequently, true superfluidity or superconductivity can be achieved with even higher critical temperature, if an ideal $I_{CF}$ configuration is designed. An important conceptual breakthrough in this letter is that the observed excitonic transport is primarily governed by the interface properties at Gr/MoS$_2$ heterostructure, rather than the intrinsic properties of each layer. Our approach provides a high feasibility for quantum dissipationless devices towards integration with available large-area CVD samples of graphene and other 2D semiconducting materials.

**Methods**

Monolayer graphene and $MoS_2$ were synthesized using a chemical vapour deposition (CVD) method. For high crystallinity, we synthesized graphene on pre-treated Cu foil, and $MoS_2$ on a $SiO_2$/Si substrate from the liquid phase.

**CVD graphene synthesis.** The high quality graphene flakes were obtained from a clean and large grain size for the Cu foil. The Cu foil was first pre-annealed at 1070℃ for 2 hours under an $H_2$ and Ar atmosphere, since the Cu grains can merge together easily near the melting point. The Cu foil was then chemically polished by a Cu etchant ($FeCl_3$, Taekwang) and rinsed with deionized (DI) water, to obtain a flat and organic residue-free substrate. Before growth, the Cu foil was annealed again at 1070 °C for 30 min to remove any residual molecules on the surface. During the CVD graphene synthesis, a small amount of low concentration $CH_4$ gas (0.1% Ar-based gas) was flowed into the chamber. The optimal condition for graphene growth was 1070℃ for 30 min, under 3 sccm $CH_4$, 20 sccm $H_2$ (99.9999%), and 1000sccm Ar gases (99.9999%).

**CVD $MoS_2$ synthesis.** Three base solutions were used; (i) OptiPrep density gradient medium (Sigma-Aldrich, D1556, 60% of (w/v) solution of iodixanol in water), (ii) sodium cholate (SC) hydrate (Sigma-Aldrich, C6445) in DI water, and (iii) solution-phase ammonium heptamolybdate (AHM) precursor (dissolving ammonium molybdate tetrahydrate (Sigma-Aldrich, 431346) into DI water). They were mixed with the ratio of (i):(ii):(iii) = 0.55:3:1. The total mixed solution was subsequently dropped onto a $SiO_2$/Si wafer, and dispersed by a spin-casting process. A two-zone CVD system (zone 1 for 200 mg of sulfur and zone 2 for substrate at atmospheric pressure) was used[23-24].

**Electrical characterizations.** The temperature-dependent static and Hall, effect measurements were performed with a standard semiconductor characterization system (B1500A, Keysight Technologies and 4200-SCS, Keithley Instruments), under a high vacuum (~$10^{-7}$ torr) in a cryostat (PPMS, Quantum Design, Inc.).

**Supplementary Information** is available in the online version of the paper.

**Acknowledgments** This work was supported by the Institute for Basic Science (IBS-R011-D1), Republic of Korea.

**Author Contributions** M.-K. Joo, Y. Jin. conceived the study and conducted the static low-temperature and magnetic transport measurement including data analysis, and authoring this manuscript as main contributors under the guidance of Y.H. Lee as a primary investigator. Y. Jin fabricated the samples. H. Kim and S. Lee synthesized the CVD monolayers of, $MoS_2$ and graphene, respectively. B.H. Moon provided comments regarding exciton superfluids and Coulomb drag. All authors participated in the scientific discussions and commented on the manuscript.

**Author Information** Reprints and permissions information is available at www.nature.com/reprints.

**Competing interests statement** The authors declare that they have no competing financial interests.

**Correspondence** Readers are welcome to comment on the online version of the paper. Correspondence and requests for materials should be addressed to M.-K. Joo (mkjoo@skku.edu) and Y.H.Lee (leeyoung@skku.edu).